\documentclass{article}
\usepackage{amsmath, amsthm}

\begin{document}

\title{\normalsize \textbf{THE MACROSCOPIC APPROACH TO EXTENDED THERMODYNAMICS WITH 14 MOMENTS, UP TO WHATEVER ORDER}}
\author{\normalsize M.C. Carrisi$^1$, S. Pennisi$^2$\\
\small $^1$ Universit\`{a} degli studi di
Cagliari, Dipartimento di Matematica ed Informatica \\
\small Via Ospedale 72,\,\ 09124 Cagliari, ITALY;\\
\small  e-mail: cristina.carrisi@tiscali.it\\[2pt]
\small $^2$ Universit\`{a} degli studi di Cagliari, Dipartimento di Matematica ed
Informatica\\
\small Via Ospedale 72,\,\ 09124 Cagliari, ITALY;\\
\small e-mail: spennisi@unica.it}
\date{}
\maketitle

\thispagestyle{empty} \noindent \textbf{Abstract:} Extended Thermodynamics is the natural
framework in which to study the physics of fluids, because it leads to symmetric
hyperbolic systems of field laws, thus assuming important properties such as finite
propagation speeds of shock waves and well posedness of the Cauchy problem. The closure
of the system of balance equations is obtained by imposing the entropy principle and that
of galilean relativity. If we take the components of the mean field as independent
variables, these two principles are equivalent to some conditions on the entropy density
and its flux. The method until now used to exploit these conditions, with the macroscopic
approach, has not been used up to whatever order with respect to thermodynamical
equilibrium. This is because it leads to several difficulties in calculations. Now these
can be overcome by using a new method proposed recently by Pennisi and Ruggeri. Here we
apply it to the 14 moments model. We will also show that the 13 moments case can be
obtained from the present one by using the method of
subsystems.\\
{\bf AMS Subject Classification:} 80A17, 74A20 \\
{\bf Key Words:} Extended Thermodynamics, entropy principle, hyperbolic systems.
\begin{center} \textbf{1. Introduction} \end{center}
The 14 moments model was firstly investigated by Kremer [1], up to second order with
respect to equilibrium; here we want to exploit it up to whatever order. The appropriate
balance equations for this model reads
\begin{eqnarray}\label{1}
&&\partial_t F+\partial_k F^k=0\nonumber\\
&&\partial_t F^i+\partial_k F^{ik}=0 \nonumber \\
&&\partial_t F^{ij}+\partial_k F^{ijk}=P^{<ij>} \nonumber\\
&&\partial_t F^{ill}+\partial_k F^{illk}=P^{ill} \nonumber\\
&&\partial_t F^{iill}+\partial_k F^{iillk}=P^{iill},
\end{eqnarray}
where the independent variables are $F$, $F^i$, $F^{ij}$, $F^{ill}$, $F^{iill}$, which
are symmetric tensors. See also ref. [2] for further details. The right hand sides of
eqs. $(\ref{1})_{1,2}$ are zero, such as the trace of that that in eq. $(\ref{1})_3$ for
the conservation laws of mass, momentum and energy. The entropy principle for these
equations, by using Liu's theorem [3], ensures the existence of parameters called
Lagrange multipliers, or mean field, such that
\begin{eqnarray}\label{2}
&&  dh=\lambda dF+\lambda_i dF^i+\lambda_{ij}dF^{ij}+\lambda_{ill}dF^{ill}+\lambda_{iill}dF^{iill}\nonumber\\
&&  d\phi^k=\lambda dF^k+\lambda_i dF^{ki}+\lambda_{ij}dF^{kij}+\lambda_{ill}dF^{kill}+\lambda_{iill}dF^{kiill}\nonumber\\
&&  \sigma=\lambda_{ij}P^{<ij>}+\lambda_{ill}P^{ill}+\lambda_{iill}P^{iill}\geq 0.
\end{eqnarray}
Following the idea exposed in ref. [4], we take the components of the mean field as
independent variables and define
\begin{eqnarray}\label{3}
&& h'=\lambda F+\lambda_i F^i+\lambda_{ij}F^{ij}+\lambda_{ill}F^{ill}+\lambda_{iill}F^{iill}-h\nonumber\\
&& \phi^{'k}=\lambda F^k+\lambda_i
F^{ki}+\lambda_{ij}F^{kij}+\lambda_{ill}F^{kill}+\lambda_{iill}F^{kiill}-\phi^k.
\end{eqnarray}
By differentiating eqs. $(\ref{3})$ and using eqs. $(\ref{2})_{1,2}$ we obtain
\begin{eqnarray}\label{4}
&&  dh'=F d\lambda +F^i d\lambda_i +F^{ij} d\lambda_{ij}+F^{ill}d\lambda_{ill}+F^{iill}d\lambda_{iill}\nonumber\\
&&  d\phi^{'k}=F^k d\lambda +F^{ki} d\lambda_i +F^{kij} d\lambda_{ij}+F^{kill}
d\lambda_{ill}+F^{kiill} d\lambda_{iill}.
\end{eqnarray}
In the next section a new methodology recently proposed by Pennisi and Ruggeri [5] will
be applied (see also [6]) to investigate eqs. $(\ref{4})$ together to those expressing
the Galilean Relativity principle, showing that they are equivalent to the subsequent
conditions $(\ref{10})$, $(\ref{13})$ and $(\ref{15})$. The last one of these will be
investigated in section 3 while the other two in section 4. To this end we will need the
expansion of $h'$ and $\phi^{'k}$ up to whatever order with respect to equilibrium;
it will be introduced also in the next section.\\
In section 5 it will be shown how the 13 moments model can be obtained as a subsystem of
the present one.\\
In section 6 we will see that the results of the kinetic approach are a particular case
of those here found with the macroscopic approach.\\
Finally conclusions will be drawn.
\begin{center} \textbf{2. The Galilean relativity principle and the entropy principle} \end{center}
We want now to impose the galilean relativity principle. To this end we recall firstly
how variables transform with a change of galileanly equivalent frames with relative
velocity $\underline{v}$. For the independent variables, from refs. [1], [2], [7] we have
\begin{eqnarray}\label{4a}
F&=&m \nonumber\\
F_i&=&m_i+m v_i \nonumber\\
F_{ij}&=&m_{ij}+2m_{(i} v_{j)}+mv_iv_j \nonumber\\
F_{ill}&=&m_{ill}+m_{ll} v_{i}+2m_{il}v_l+m_iv^2+2m_lv_lv_i+mv^2v_i \nonumber\\
F_{iill}&=&m_{iill}+4m_{ill} v_{i}+2m_{ll}v^2+4m_{li}v_iv_l+4m_lv_lv^2+mv^4;
\end{eqnarray}
here the $m_{...}$ are the tensors corresponding to $F_{...}$ in the second reference
frame. Moreover we have
\begin{eqnarray}\label{4b}
F_k&=&F v_k+m_k\nonumber\\
F_{ik}&=&F_iv_k+m_{ik}+m_{k} v_{i}\nonumber\\
F_{ijk}&=&F_{ij}v_k+m_{ijk}+2m_{k(i} v_{j)}+m_kv_iv_j\nonumber\\
F_{illk}&=&F_{ill}v_k+m_{illk}+m_{kll} v_{i}+2m_{kil}v_l+m_{ki}v^2+2m_{kl}v_lv_i+m_kv^2v_i\nonumber\\
F_{iillk}&=&F_{iill}v_k+m_{iillk}+4m_{kill}
v_{i}+2m_{kll}v^2+4m_{kli}v_lv_i+4m_{kl}v_lv^2+m_kv^4\nonumber \\
h&=&\hat{h}\nonumber\\
\phi^k&=&\hat{h}v_k+\hat{\phi}^k.
\end{eqnarray}
The first two of these, as the trace of the third and fourth ones, are identities, while
what remains is the transformation law of the dependent variables. Substituting the
relations above into eq. $(\ref{2})_{1}$ and defining
\begin{eqnarray}\label{9}
&&  \hat{\lambda}=\lambda+\lambda_iv_i+\lambda_{ij}v_i
  v_j+\lambda_{ipp}v^2v_i+\lambda_{ppqq}v^4 \nonumber \\
&&\hat{\lambda}_i=\lambda_i+2\lambda_{ij} v_j+2\lambda_{jpp}v_jv_i+\lambda_{ipp}v^2+4\lambda_{ppqq}v^2v_i \nonumber \\
&&\hat{\lambda}_{ij}=\lambda_{ij}+\lambda_{hpp}v_h\delta_i^j+2\lambda_{ipp}v_j+2\lambda_{ppqq}v^2\delta_i^j+4\lambda_{ppqq}v_iv_j \nonumber \\
&&\hat{\lambda}_{ill}=\lambda_{ipp}+4\lambda_{ppqq}v_i \nonumber \\
&&\hat{\lambda}_{ppqq}=\lambda_{ppqq}
\end{eqnarray}
we have
\begin{equation}\label{6}
d\hat{h}=\hat{\lambda}dm+\hat{\lambda}_i
dm_i+\hat{\lambda}_{ij}dm_{ij}+\hat{\lambda}_{ill}dm_{ill}+\hat{\lambda}_{iill}dm_{iill}.
\end{equation}
For eq. $(\ref{4b})_6$ we note that eq. $(\ref{6})$ is the counterpart of eq.
$(\ref{2})_1$ in the second frame; this allows us to see that eqs. $(\ref{9})$ are the
transformation rules for the Lagrange Multipliers.\\
Similarly, by substituting eqs. $(\ref{4b})$ in eq. $(\ref{2})_2$, we find
\begin{equation}\label{6a}
d\hat{\phi}^k=\hat{\lambda}dm_k+\hat{\lambda}_i
dm_{ki}+\hat{\lambda}_{ij}dm_{kij}+\hat{\lambda}_{ill}dm_{kill}+\hat{\lambda}_{iill}dm_{kiill}
\end{equation}
which is the counterpart of eq. $(\ref{2})_2$ in other frame.\\
The counterparts of eqs. $(\ref{3})$ in the second frame are
\begin{eqnarray}\label{6b}
\hat{h'}&=&m\hat{\lambda}+ m_i\hat{\lambda}_{i}+m_{ij} \hat{\lambda}_{ij}+m_{ill}
\hat{\lambda}_{ill}+m_{iill}\hat{\lambda}_{iill}-\hat{h}\nonumber \\
\hat{\phi}^{'k}&=&m_k\hat{\lambda}+ m_{ki}\hat{\lambda}_{i}+m_{kij}
\hat{\lambda}_{ij}+m_{kill}
\hat{\lambda}_{ill}+m_{kiill}\hat{\lambda}_{iill}-\hat{\phi}^k;
\end{eqnarray}
differentiating them and using eqs. $(\ref{4b})_{6,7}$, $(\ref{6})$ and $(\ref{6a})$ we
obtain respectively
\begin{eqnarray*}
d\hat{h'}&=&m d\hat{\lambda}+ m_id\hat{\lambda}_{i}+m_{ij} d\hat{\lambda}_{ij}+m_{ill}
d\hat{\lambda}_{ill}+m_{iill}d\hat{\lambda}_{iill},\\
d\hat{\phi}^{'k}&=&m_k d\hat{\lambda}+ m_{ki}d\hat{\lambda}_{i}+m_{kij}
d\hat{\lambda}_{ij}+m_{kill} d\hat{\lambda}_{ill}+m_{kiill}d\hat{\lambda}_{iill}.
\end{eqnarray*}
Taking their derivatives with respect to the various components of the main field we have
\begin{eqnarray*}\label{11}
m=\frac{\partial \hat{h}'}{\partial \hat{\lambda}}, \qquad m_i=\frac{\partial
\hat{h}'}{\partial \hat{\lambda}_i}, \qquad m_{ij}=\frac{\partial \hat{h}'}{\partial
\hat{\lambda}_{ij}}, \quad m_{ill}=\frac{\partial \hat{h}'}{\partial
\hat{\lambda}_{ill}}, \qquad m_{iill}=\frac{\partial \hat{h}'}{\partial
\hat{\lambda}_{iill}}, \\
m_k=\frac{\partial \hat{\phi}^{'k}}{\partial \hat{\lambda}}, \quad m_{ki}=\frac{\partial
\hat{\phi}^{'k}}{\partial \hat{\lambda}_i}, \quad m_{kij}=\frac{\partial
\hat{\phi}^{'k}}{\partial \hat{\lambda}_{ij}}, \quad m_{kill}=\frac{\partial
\hat{\phi}^{'k}}{\partial \hat{\lambda}_{ill}}, \quad m_{kiill}=\frac{\partial
\hat{\phi}^{'k}}{\partial \hat{\lambda}_{iill}}. \\
(11)
\end{eqnarray*}
Comparing the correspondent terms in the two rows of eq. $(\ref{11})$ we obtain the
following compatibility conditions: \setcounter{equation}{11}
\begin{eqnarray}\label{15}
&&\frac{\partial \hat{h}'}{\partial \hat{\lambda}_k}=\frac{\partial
\hat{\phi}^{'k}}{\partial \hat{\lambda}}, \qquad \frac{\partial \hat{h}'}{\partial
\hat{\lambda}_{ki}}=\frac{\partial \hat{\phi}^{'k}}{\partial \hat{\lambda}_i}, \qquad
\frac{\partial \hat{h}'}{\partial \hat{\lambda}_{ill}}=\frac{\partial
\hat{\phi}^{'k}}{\partial \hat{\lambda}_{ij}}\delta_i^j, \nonumber \\
&&\frac{\hat{\phi}^{'k]}}{\partial \hat{\lambda}_{i[j}}=0, \qquad \frac{\partial
\hat{h}'}{\partial \hat{\lambda}_{kkll}}=\frac{\partial \hat{\phi}^{'k}}{\partial
\hat{\lambda}_{ill}}\delta_i^k, \qquad \frac{\partial \hat{\phi}^{'k]}}{\partial
\hat{\lambda}_{ll[i}}=0.
\end{eqnarray}
By substituting $h$, $\hat{h}$, $\phi^k$ and $\hat{\phi}^k$ from eqs. $(\ref{3})_1$,
$(\ref{6b})_1$, $(\ref{3})_2$, $(\ref{6b})_2$ into eqs. $(\ref{4b})_{6,7}$, these become
\begin{equation}\label{11a}
h'=\hat{h}', \quad \quad \phi^{'k}=\hat{\phi}^{'k}+\hat{h}'v^k,
\end{equation}
where $(\ref{4b})_{1-5}$ and $(\ref{9})$ have been used.\\
Now, from eqs. $(\ref{11a})$ we see that $h'$ and $\phi^{'k}$ are composite functions of
$\hat{h}'$ and $\hat{\phi}^{'k}$ and of eqs. $(\ref{9})$; but $h'$ and $\phi^{'k}$ depend
only on $\lambda$, $\lambda_{i}$, $\lambda_{ij}$, $\lambda_{ill}$, $\lambda_{iill}$ and
not on $v_h$. In other words, the derivative of $h'$ and $\phi^{'k}$ with respect to
$v_h$, through the above mentioned composite functions, must be zero, i.e.
\begin{equation}\label{10}
\frac{\partial h'}{\partial v_h}=0=m\hat{\lambda}_h+ 2m_i\hat{\lambda}_{ih}+
\hat{\lambda}_{ipp}\left(m_{ll}\delta_i^h+2m_{ih}\right)+4m_{hll}\hat{\lambda}_{ppqq}~~~~~~~~~~~~
\end{equation}
\begin{equation}\label{13}
\frac{\partial \phi^{'k}}{\partial v_h}=0=m_k\hat{\lambda}_h+ 2m_{ki}\hat{\lambda}_{ih}+
\hat{\lambda}_{ipp}\left(m_{kll}\delta_i^h+2m_{kih}\right)+4m_{khll}\hat{\lambda}_{ppqq}+\delta_h^kh'
\end{equation}
where $(\ref{11})$ and $(\ref{9})$ have been used.\\
The entropy principle and that of material objectivity reduce in imposing eqs.
$(\ref{10})$, $(\ref{13})$ and $(\ref{15})$. We want to impose these conditions up to
whatever order with respect to thermodynamical equilibrium. This is defined, see [8], as
the state where all the components of the main field, except $\hat{\lambda}$ and
$\hat{\lambda}_{ij}=\frac{1}{3}\hat{\lambda}_{ll}\delta_{ij}$, amounts to zero. To avoid
an excessive quantity of indexes, we will do later the expansion with respect to
$\hat{\lambda}_{ppqq}$. The expansion of the tensor $\hat{\phi}^{'i}$ with respect to the
other variables is
\begin{eqnarray}\label{16}
\hat{\phi}^{'i}=&&\sum_{p=0}^{\infty}\sum_{q=0}^{\infty}\sum_{r=0}^{\infty}\frac{1}{p!q!r!}
\phi_{p,q,r}^{ii_1\cdots i_pj_1\cdots j_q k_1h_1\cdots k_r h_r}\hat{\lambda}_{i_1}\cdots
\hat{\lambda}_{i_p}\hat{\lambda}_{j_1ll}\cdots \hat{\lambda}_{j_qll}\cdot \nonumber \\
&&\left(\hat{\lambda}_{k_1h_1}-\frac{1}{3}\hat{\lambda}_{ll}\delta_{k_1h_1}\right) \cdots
\left(\hat{\lambda}_{k_rh_r}-\frac{1}{3}\hat{\lambda}_{ll}\delta_{k_rh_r}\right)
\end{eqnarray}
\begin{eqnarray}\label{17} \text{with }&&\phi_{p,q,r}^{ii_1\cdots i_pj_1\cdots j_q k_1h_1\cdots k_r
h_r}(\hat{\lambda},\hat{\lambda}_{ll},\hat{\lambda}_{ppqq})=\nonumber \\
&&=\left(\frac{\partial^{p+q+r} \hat{\phi}^{'i}}{\partial
\hat{\lambda}_{i_1}\cdots\partial\hat{\lambda}_{i_p}\partial\hat{\lambda}_{j_1ll}\cdots\partial
\hat{\lambda}_{j_qll}\partial\hat{\lambda}_{k_1h_1}\cdots
\partial\hat{\lambda}_{h_rk_r}}\right)_{eq}.~~~~~~~~~~
\end{eqnarray}
Now, from the compatibility conditions $(\ref{15})_2$, $(\ref{15})_6$ and $(\ref{15})_4$
we see that we can exchange the index i respectively with each other index taken from
$i_1\cdots i_p$, $j_1,\cdots j_q$ and $h_1\cdots h_r$ or $k_1\cdots k_r$, so
$\phi_{p,q,r}^{ii_1\cdots i_pj_1\cdots j_q k_1h_1\cdots k_r h_r}$ is a symmetric tensor
with respect to any couple of indexes. Moreover $\phi_{p,q,r}^{ii_1\cdots i_pj_1\cdots
j_q k_1h_1\cdots k_r h_r}$ depends only on scalars, so that
\begin{equation}\label{18}
  \begin{cases}
   \phi_{p,q,r}^{ii_1\cdots i_pj_1\cdots j_q k_1h_1\cdots k_r h_r}=0 & \text{if p+q+2r+1 is odd}\\
   \phi_{p,q,r}^{ii_1\cdots i_pj_1\cdots j_q k_1h_1\cdots k_r h_r}=\phi_{p,q,r}(\hat{\lambda},\hat{\lambda}_{ll},\hat{\lambda}_{ppqq})\delta^{ii_1}\cdots \delta^{k_r h_r} & \text{if p+q+2r+1 is
   even},
  \end{cases}
\end{equation}
so that $\phi_{p,q,r}^{ii_1\cdots i_pj_1\cdots j_q k_1h_1\cdots k_r h_r}$ is known except
for a scalar function.\\
Similarly, for the tensor $\hat{h}'$ we have
\begin{eqnarray}\label{19}
\hat{h}'=&&\sum_{p=0}^{\infty}\sum_{q=0}^{\infty}\sum_{r=0}^{\infty}\frac{1}{p!q!r!}
h_{p,q,r}^{i_1\cdots i_pj_1\cdots j_q k_1h_1\cdots k_r h_r}\hat{\lambda}_{i_1}\cdots
\hat{\lambda}_{i_p}\hat{\lambda}_{j_1ll}\cdots \hat{\lambda}_{j_qll}\cdot \nonumber \\
&&\left(\hat{\lambda}_{k_1h_1}-\frac{1}{3}\hat{\lambda}_{ll}\delta_{k_1h_1}\right) \cdots
\left(\hat{\lambda}_{k_rh_r}-\frac{1}{3}\hat{\lambda}_{ll}\delta_{k_rh_r}\right)
\end{eqnarray}
\begin{eqnarray}\label{20} \text{with }&&h_{p,q,r}^{i_1\cdots i_pj_1\cdots j_q k_1h_1\cdots k_r
h_r}(\hat{\lambda},\hat{\lambda}_{ll},\hat{\lambda}_{ppqq})=\nonumber \\
&&= \left(\frac{\partial^{p+q+r} \hat{h}'}{\partial
\hat{\lambda}_{i_1}\cdots\partial\hat{\lambda}_{i_p}\partial\hat{\lambda}_{j_1ll}\cdots\partial
\hat{\lambda}_{j_qll}\partial\hat{\lambda}_{k_1h_1}\cdots
\partial\hat{\lambda}_{h_rk_r}}\right)_{eq}.~~~~~~
\end{eqnarray}
Taking the derivatives with respect to $\hat{\lambda}_{jll}$ of the compatibility
conditions $(\ref{15})_1$ and $(\ref{15})_2$ and using $(\ref{15})_6$ we see that we can
exchange every index taken from $j_1,\cdots j_q$ with each other. Similarly, taking the
derivative of eq. $(\ref{15})_1$ with respect to $\hat{\lambda}_{rs}$ and using eq.
$(\ref{15})_4$ we see that we can exchange every index taken from $i_1,\cdots, i_p$ with
each other. Consequently, $h_{p,q,r}^{i_1\cdots i_pj_1\cdots j_q k_1h_1\cdots k_r h_r}$
is a symmetric tensor with respect to any couple of indexes; moreover it depends only on
scalars, so that
\begin{equation}\label{21}
  \begin{cases}
   h_{p,q,r}^{i_1\cdots i_pj_1\cdots j_q k_1h_1\cdots k_r h_r}=0 &\text{if p+q+2r is odd}\\
   h_{p,q,r}^{i_1\cdots i_pj_1\cdots j_q k_1h_1\cdots k_r h_r}=h_{p,q,r}(\hat{\lambda},\hat{\lambda}_{ll},\hat{\lambda}_{ppqq})\delta^{i_1i_2}\cdots \delta^{k_r h_r}&\text{if p+q+2r is
   even}.
  \end{cases}
\end{equation}
In other words, also $h_{p,q,r}^{i_1\cdots i_pj_1\cdots j_q k_1h_1\cdots k_r h_r}$ is
known except for a scalar function.\\
We want to avoid to use eqs. $(\ref{17})$ and $(\ref{20})$ in the sequel. To this end we
note that we can consider
\begin{equation*}
\frac{\partial^{p+q+r} \hat{h}'}{\partial
\hat{\lambda}_{i_1}\cdots\partial\hat{\lambda}_{i_p}\partial\hat{\lambda}_{j_1ll}\cdots\partial
\hat{\lambda}_{j_qll}\partial\hat{\lambda}_{k_1h_1}\cdots
\partial\hat{\lambda}_{h_rk_r}}
\end{equation*}
\begin{equation*}
\text{and } \quad \frac{\partial^{p+q+r} \hat{\phi}^{'k}}{\partial
\hat{\lambda}_{i_1}\cdots\partial\hat{\lambda}_{i_p}\partial\hat{\lambda}_{j_1ll}\cdots\partial
\hat{\lambda}_{j_qll}\partial\hat{\lambda}_{k_1h_1}\cdots
\partial\hat{\lambda}_{h_rk_r}}\qquad
\end{equation*}
depending on $\hat{\lambda}_{ab}$ as composite functions through
$\hat{\lambda}_{<ab>}=\left(\delta_a^i
\delta_b^j-\frac{1}{3}\delta^{ij}\delta_{ab}\right)\hat{\lambda}_{ij}$ and
$\hat{\lambda}_{ll}$. With this in mind let us take their derivatives with respect to
$\hat{\lambda}_{ab}$, after that contract them with $\delta_{ab}$ and calculate the
result at equilibrium; we find
\begin{equation}\label{22}
h_{p,q,r+1}^{i_1\cdots i_pj_1\cdots j_q k_1h_1\cdots k_r h_r
ab}\delta_{ab}=3\frac{\partial}{\partial \hat{\lambda}_{ll}}h_{p,q,r}^{i_1\cdots
i_pj_1\cdots j_q k_1h_1\cdots k_r h_r}
\end{equation}
\begin{equation}\label{23}
\text{and } \quad \phi_{p,q,r+1}^{ii_1\cdots i_pj_1\cdots j_q k_1h_1\cdots k_r h_r
ab}\delta_{ab}=3\frac{\partial}{\partial \hat{\lambda}_{ll}}\phi_{p,q,r}^{ii_1\cdots
i_pj_1\cdots j_q k_1h_1\cdots k_r h_r}.
\end{equation}
An interesting consequence of eq. $(\ref{22})$ can be observed as follows.\\
Let us take the derivative of $\hat{h}'$ with respect to $\hat{\lambda}_{ij}$ taking into
account that
$\hat{\lambda}_{ij}=\frac{1}{3}\hat{\lambda}_{ll}\delta_{ij}+\hat{\lambda}_{<ij>}$:\newpage
\begin{eqnarray*}
&&\frac{\partial \hat{h}'}{\partial
\hat{\lambda}_{ij}}=\sum_{p=0}^{\infty}\sum_{q=0}^{\infty}\sum_{r=0}^{\infty}\frac{1}{p!q!r!}
\frac{\partial h_{p,q,r}^{i_1\cdots i_pj_1\cdots j_q k_1h_1\cdots k_r h_r}}{\partial
\hat{\lambda}_{ll}}\hat{\lambda}_{i_1}\cdots
\hat{\lambda}_{i_p}\hat{\lambda}_{j_1ll}\cdots \hat{\lambda}_{j_qll}\cdot \\
&&\left(\hat{\lambda}_{k_1h_1}-\frac{1}{3}\hat{\lambda}_{ll}\delta_{k_1h_1}\right) \cdots
\left(\hat{\lambda}_{k_rh_r}-\frac{1}{3}\hat{\lambda}_{ll}\delta_{k_rh_r}\right)\delta_{ij}+\\
&&+\sum_{p=0}^{\infty}\sum_{q=0}^{\infty}\sum_{r=1}^{\infty}\frac{r}{p!q!r!}
h_{p,q,r}^{i_1\cdots i_pj_1\cdots j_q k_1h_1\cdots k_r h_r}\hat{\lambda}_{i_1}\cdots
\hat{\lambda}_{i_p}\hat{\lambda}_{j_1ll}\cdots \hat{\lambda}_{j_qll}\cdot  \\
&&\left(\hat{\lambda}_{k_1h_1}-\frac{1}{3}\hat{\lambda}_{ll}\delta_{k_1h_1}\right) \cdots
\left(\hat{\lambda}_{k_{r-1}h_{r-1}}-\frac{1}{3}\hat{\lambda}_{ll}\delta_{k_{r-1}h_{r-1}}\right)
\left(\delta_{h_r}^i\delta_{k_r}^j-\frac{1}{3}\delta_{h_rk_r}\delta^{ij}\right),
\end{eqnarray*}
which, by using eq. $(\ref{22})$, becomes
\begin{eqnarray*}
\frac{\partial \hat{h}'}{\partial \hat{\lambda}_{ij}}
&=&\Bigg[\sum_{p=0}^{\infty}\sum_{q=0}^{\infty}\sum_{r=0}^{\infty}\frac{1}{p!q!r!}\frac{1}{3}
h_{p,q,r+1}^{i_1\cdots i_pj_1\cdots j_q k_1h_1\cdots k_r h_r ab}\delta_{ab}
\hat{\lambda}_{i_1}\cdots
\hat{\lambda}_{i_p}\cdot \\
&&\hat{\lambda}_{j_1ll}\cdots
\hat{\lambda}_{j_qll}\left(\hat{\lambda}_{k_1h_1}-\frac{1}{3}\hat{\lambda}_{ll}\delta_{k_1h_1}\right)
\cdots
\left(\hat{\lambda}_{k_rh_r}-\frac{1}{3}\hat{\lambda}_{ll}\delta_{k_rh_r}\right)\delta^{ij}+\\
&-&\sum_{p=0}^{\infty}\sum_{q=0}^{\infty}\sum_{r=1}^{\infty}\frac{r}{p!q!r!}\frac{1}{3}
h_{p,q,r}^{i_1\cdots i_pj_1\cdots j_q k_1h_1\cdots k_r h_r}\hat{\lambda}_{i_1}\cdots
\hat{\lambda}_{i_p}\hat{\lambda}_{j_1ll}\cdots \hat{\lambda}_{j_qll}\cdot  \\
&&\left(\hat{\lambda}_{k_1h_1}-\frac{1}{3}\hat{\lambda}_{ll}\delta_{k_1h_1}\right) \cdots
\left(\hat{\lambda}_{k_{r-1}h_{r-1}}-\frac{1}{3}\hat{\lambda}_{ll}\delta_{k_{r-1}h_{r-1}}\right)
\delta_{h_rk_r}\Bigg]\delta^{ij}+\\
&+&\sum_{p=0}^{\infty}\sum_{q=0}^{\infty}\sum_{r=1}^{\infty}\frac{r}{p!q!r!}
h_{p,q,r}^{i_1\cdots i_pj_1\cdots j_q k_1h_1\cdots k_r h_r}\hat{\lambda}_{i_1}\cdots
\hat{\lambda}_{i_p}\hat{\lambda}_{j_1ll}\cdots \hat{\lambda}_{j_qll}\cdot  \\
&&\left(\hat{\lambda}_{k_1h_1}-\frac{1}{3}\hat{\lambda}_{ll}\delta_{k_1h_1}\right) \cdots
\left(\hat{\lambda}_{k_{r-1}h_{r-1}}-\frac{1}{3}\hat{\lambda}_{ll}\delta_{k_{r-1}h_{r-1}}\right)
\delta_{h_r}^i\delta^{j}_{k_r};
\end{eqnarray*}
We note that the term in square brackets amounts to zero as can be easily proved by
substituting r=R+1 in the second sum. What remains can be written as
\begin{equation*}
\frac{\partial \hat{h}'}{\partial \hat{\lambda}_{ij}}=\frac{\partial \hat{h}'}{\partial
\hat{\lambda}_{<ij>}},
\end{equation*}
where the derivative in the right hand side has been taken without considering that the
components of $\hat{\lambda}_{<ij>}$ aren't independent because restricted by
$\hat{\lambda}_{<ij>}\delta^{ij}=0$. Proceeding similarly with $\hat{\phi}^{'k}$ and
using eq. $(\ref{23})$ we find that
\begin{equation*}
\frac{\partial \hat{\phi}^{'k}}{\partial \hat{\lambda}_{ij}}=\frac{\partial
\hat{\phi}^{'k}}{\partial \hat{\lambda}_{<ij>}}.
\end{equation*}
After that, we see that eq. $(\ref{17})$ and $(\ref{20})$ become consequences of eqs.
$(\ref{16})$ and $(\ref{19})$ so that they can be forgotten. But, instead of them, we
have to impose eqs. $(\ref{22})$ and $(\ref{23})$. \\
Expliciting eq. $(\ref{22})$ by means of eq. $(\ref{21})$ we have
\begin{equation}\label{24}
h_{p,q,r+1}=3\frac{p+q+2r+1}{p+q+2r+3}\frac{\partial h_{p,q,r}}{\partial
\hat{\lambda}_{ll}},
\end{equation}
from which
\begin{equation}\label{25}
h_{p,q,r}=3^r\frac{p+q+1}{p+q+2r+1}\frac{\partial^r h_{p,q,0}}{\partial
\hat{\lambda}_{ll}^r},
\end{equation}
as it can be seen by using the iterative procedure.\\
Similarly, expliciting eq. $(\ref{23})$ by means of eq. $(\ref{18})$, we have
\begin{equation}\label{26}
\phi_{p,q,r+1}=3\frac{p+q+2r+2}{p+q+2r+4}\frac{\partial \phi_{p,q,r}}{\partial
\hat{\lambda}_{ll}},
\end{equation}
from which
\begin{equation}\label{27}
\phi_{p,q,r}=3^r\frac{p+q+2}{p+q+2r+2}\frac{\partial^r \phi_{p,q,0}}{\partial
\hat{\lambda}^{r}_{ll}},
\end{equation}
that can be proved using the iterative procedure.\\
If we introduce the quantities
\begin{equation}\label{definition}
  \begin{cases}
    k_{p,q}=h_{p,q,0} \quad \text{if p+q is even}\\
    k_{p,q}=\phi_{p,q,0} \quad \text{if p+q is odd},
  \end{cases}
\end{equation}
we note that $\hat{h}'$ and $\hat{\phi}^{'k}$ are known if we know all the terms of the
infinity matrix $k_{p,q}$; so our aim is to find $k_{p,q}$. We have also to impose the
compatibility conditions $(\ref{15})$ and the conditions $(\ref{10})$ and $(\ref{13})$
expressing the Galilean relativity principle. Let us begin by investigating the
conditions $(\ref{15})$.
\begin{center} \textbf{3. Exploitation of the conditions $(\ref{15})$} \end{center}
Now let's impose conditions $(\ref{15})$ on our tensors. We notice that equations
$(\ref{15})_{4,6}$ are already satisfied because the tensors $\phi_{p,q,r}^{ii_1\cdots
i_p j_1\cdots j_q h_1k_1 \cdots h_r k_r}$ are symmetric, so there remains to impose eqs.
$(\ref{15})_{1,2,3,5}$. \\~\\
$\bullet$~~ Eq. $(\ref{15})_1$, by using $(\ref{16})$, $(\ref{18})$, $(\ref{20})$ and
$(\ref{21})$, becomes
\begin{equation}\label{alfa}
h_{p+1,q,r}=\frac{\partial \phi_{p,q,r}}{\partial \hat{\lambda}}
\end{equation}
which, for r=0 reads
\begin{equation}\label{alfa1}
h_{p+1,q,0}=\frac{\partial \phi_{p,q,0}}{\partial \hat{\lambda}}
\end{equation}
and, for the other values of r is consequence of $(\ref{25})$, $(\ref{27})$,
$(\ref{alfa1})$. This last one, by using $(\ref{definition})$, can be written also as
\begin{equation}\label{beta1}
k_{p+1,q}=\frac{\partial k_{p,q}}{\partial \hat{\lambda}} \quad \text{with p+q+1 even}.
\end{equation}
In other words, the elements with p+q+1 even of the matrix $k_{p+1,q}$ can be expressed
in terms of that of the same column but previous row. \\~~~\\
$\bullet$ ~~Let us impose now eq. $(\ref{15})_2$, using eqs. $(\ref{16})$, $(\ref{18})$,
$(\ref{20})$ and $(\ref{21})$; we obtain
\begin{equation}\label{gamma}
h_{p,q,r+1}=\phi_{p+1,q,r}
\end{equation}
which, by using eqs. $(\ref{25})$ and $(\ref{27})$ is equivalent to
\begin{equation}\label{gamma*}
\phi_{p+1,q,0}=3\frac{p+q+1}{p+q+3}\frac{\partial h_{p,q,0}}{\partial \hat{\lambda}_{ll}}
\end{equation}
and this, by using $(\ref{definition})$, becomes
\begin{equation}\label{gamma1}
k_{p+1,q}=3\frac{p+q+1}{p+q+3}\frac{\partial k_{p,q}}{\partial \hat{\lambda}_{ll}} \quad
\text{with p+q even}.
\end{equation}
Using $(\ref{beta1})$ or $(\ref{gamma1})$ we can express all the elements of the matrix
$k_{p,q}$ in terms of those in the same column and previous row. Iterating this procedure
each element can be expressed in terms of the elements in the first row of the matrix. In
fact joining eqs. $(\ref{beta1})$ and $(\ref{gamma1})$ we obtain
\begin{equation}\label{betagamma}
  \begin{cases}
    k_{p,q}=3^{\frac{p}{2}}\frac{q+1}{p+q+1}\frac{\partial^p}{\partial \hat{\lambda}_{ll}ì^{\frac{p}{2}}
    \partial\hat{\lambda}^{\frac{p}{2}}} k_{0,q} & \text{with p and q even}, \\
    k_{p,q}=3^{\frac{p-1}{2}}\frac{q+2}{p+q+1}\frac{\partial^p}{\partial \hat{\lambda}_{ll}ì^{\frac{p-1}{2}}
    \partial\hat{\lambda}^{\frac{p+1}{2}}}k_{0,q} & \text{with p and q odd}, \\
    k_{p,q}=3^{\frac{p}{2}}\frac{q+2}{p+q+2}\frac{\partial^p}{\partial \hat{\lambda}_{ll}ì^{\frac{p}{2}}
    \partial\hat{\lambda}^{\frac{p}{2}}}k_{0,q} & \text{with p even and q odd}, \\
    k_{p,q}=3^{\frac{p+1}{2}}\frac{q+1}{p+q+2}\frac{\partial^p}{\partial \hat{\lambda}_{ll}ì^{\frac{p+1}{2}}
    \partial\hat{\lambda}^{\frac{p-1}{2}}}k_{0,q} & \text{with p odd and q even}.
  \end{cases}
\end{equation}
$\bullet$ ~~Finally, let us consider eqs. $(\ref{15})_{3,5}$. Using eqs. $(\ref{16})$,
$(\ref{18})$, $(\ref{20})$ and $(\ref{21})$ they become res\-pec\-ti\-ve\-ly
\begin{equation}\label{delta}
h_{p,q+1,r}=\frac{p+q+2r+4}{p+q+2r+2}\phi_{p,q,r+1}
\end{equation}
and
\begin{equation}\label{epsilon}
\frac{\partial
h_{p,q,r}}{\partial\hat{\lambda}_{kkll}}=\frac{p+q+2r+3}{p+q+2r+1}\phi_{p,q+1,r}.
\end{equation}
By using eqs. $(\ref{25})$, $(\ref{27})$ and finally $(\ref{definition})$ the above
equations transform respectively into
\begin{equation}\label{epsilon1}
k_{p,q+1}=3\frac{\partial k_{p,q}}{\partial \hat{\lambda}_{ll}} \quad \text{with p+q+1
even}
\end{equation}
and
\begin{equation}\label{epsilon2}
k_{p,q+1}=\frac{p+q+1}{p+q+3}\frac{\partial k_{p,q}}{\partial \hat{\lambda}_{aabb}} \quad
\text{with p+q+1 odd}.
\end{equation}
In other words with eqs. $(\ref{epsilon1})$ and $(\ref{epsilon2})$ each element of the
matrix $k_{p,q}$ can be written in terms of the element in the same row and previous
column. But we already know, by eqs. $(\ref{betagamma})$, each row of the matrix
$k_{p,q}$ in terms of the first one; so we have to investigate the compatibility of these
two results. By substituting eqs. $(\ref{betagamma})$ into eqs. $(\ref{epsilon1})$ and
$(\ref{epsilon2})$ we obtain a series of equations for the first row of the matrix
$k_{p,q}$, i.e.,
\begin{equation}\label{fi}
  \begin{cases}
    k_{0,q+1}=3\frac{\partial}{\partial \hat{\lambda}_{ll}}k_{0,q} & \text{q
odd}, \\
    9\frac{q+1}{q+3}\frac{\partial^2}{\partial \hat{\lambda}_{ll}^2}k_{0,q}=\frac{\partial}{\partial \hat{\lambda}}k_{0,q+1} & \text{q even}, \\
\frac{q+1}{q+3}\frac{\partial}{\partial \hat{\lambda}_{aabb}}k_{0,q}=k_{0,q+1} & \text{q even}, \\
\frac{\partial}{\partial \hat{\lambda}_{aabb}}\frac{\partial}{\partial \hat{\lambda}
}k_{0,q}=3\frac{\partial}{\partial \hat{\lambda}_{ll}}k_{0,q+1} & \text{q odd},
  \end{cases}
\end{equation}
and other equations which are consequences of these last ones. Now eqs. $(\ref{fi})_1$
and $(\ref{fi})_3$ give each element $k_{0,q}$ in terms of $k_{0,0}$, i.e.,
\begin{equation}\label{A}
k_{0,q}=3^{\frac{q}{2}}\frac{1}{q+1}\frac{\partial^q}{\partial
\hat{\lambda}_{ll}^{\frac{q}{2}}\partial \hat{\lambda}_{aabb}^{\frac{q}{2}}}k_{0,0}\quad
\text{with q even }
\end{equation}
\begin{equation}\label{B}
k_{0,q}=3^{\frac{q-1}{2}}\frac{1}{q+2}\frac{\partial^q}{\partial
\hat{\lambda}_{ll}^{\frac{q-1}{2}}\partial
\hat{\lambda}_{aabb}^{\frac{q+1}{2}}}k_{0,0}\quad \text{with q odd}.
\end{equation}
Through these two equations is possible to express a generic element in the first row in
terms of the first element in the same first row of the matrix.\\
Eq. $(\ref{fi})_{2,4}$ remain to be imposed. The first one of these with q=0 and by use
of $(\ref{A})$ reads
\begin{equation}\label{C}
9\frac{\partial^2}{\partial \hat{\lambda}_{ll}^{2}}k_{0,0}=\frac{\partial}{\partial
\hat{\lambda}}\frac{\partial}{\partial \hat{\lambda}_{aabb}}k_{0,0},
\end{equation}
which is a condition on $k_{0,0}$. After that eq. $(\ref{fi})_2$ for the other values of q
is a consequence of eq. $(\ref{C})$. \\
At last, eq. $(\ref{fi})_4$ with use of $(\ref{A})$ and $(\ref{B})$ becomes equivalent to
its value for q=1, i.e.,
\begin{equation*}
9\frac{\partial^3}{\partial \hat{\lambda}_{ll}^{2}\partial
\hat{\lambda}_{aabb}}k_{0,0}=\frac{\partial^3}{\partial \hat{\lambda}\partial
\hat{\lambda}^2_{aabb}}k_{0,0},
\end{equation*}
which is eq. $(\ref{C})$ differentiated with respect to $\hat{\lambda}_{aabb}$; so it is
sufficient to impose eq. $(\ref{C})$.\\
We can now substitute eqs. $(\ref{A})$ and $(\ref{B})$ into eqs. $(\ref{betagamma})$
which now become
\begin{equation}\label{40bis}
  \begin{cases}
    k_{p,q}=3^{\frac{p+q}{2}}\frac{1}{p+q+1}\frac{\partial^{p+q}}{\partial \hat{\lambda}_{ll}ì^{\frac{p+q}{2}}
    \partial\hat{\lambda}^{\frac{p}{2}}\partial \hat{\lambda}_{aabb}ì^{\frac{q}{2}}} k_{0,0} & \text{with p and q even}, \\
    k_{p,q}=3^{\frac{p+q-2}{2}}\frac{1}{p+q+1}\frac{\partial^{p+q}}{\partial \hat{\lambda}_{ll}ì^{\frac{p+q-2}{2}}
    \partial\hat{\lambda}^{\frac{p+1}{2}}\partial\hat{\lambda}_{aabb}ì^{\frac{q+1}{2}}}k_{0,0} & \text{with p and q odd}, \\
    k_{p,q}=3^{\frac{p+q-1}{2}}\frac{1}{p+q+2}\frac{\partial^{p+q}}{\partial \hat{\lambda}_{ll}ì^{\frac{p+q-1}{2}}
    \partial\hat{\lambda}^{\frac{p}{2}}\partial \hat{\lambda}_{aabb}ì^{\frac{q+1}{2}}}k_{0,0} & \text{with p even and q odd}, \\
    k_{p,q}=3^{\frac{p+q+1}{2}}\frac{1}{p+q+2}\frac{\partial^{p+q}}{\partial \hat{\lambda}_{ll}ì^{\frac{p+q+1}{2}}
    \partial\hat{\lambda}^{\frac{p-1}{2}}\partial \hat{\lambda}_{aabb}ì^{\frac{q}{2}}}k_{0,0} & \text{with p odd and q even}.
  \end{cases}
\end{equation}
In this way all the elements of the matrix $k_{p,q}$ are determined in terms of $k_{0,0}$
which is restricted, until now, only by eq. $(\ref{C})$. Another restriction will be
found in the next section.
\begin{center} \textbf{4. Exploitation of the conditions $(\ref{10})$ and $(\ref{13})$} \end{center}
There remains now to impose eqs. $(\ref{10})$ and $(\ref{13})$, but we can see that
$(\ref{10})$ is a consequence of $(\ref{13})$ and $(\ref{15})$. In fact
\begin{itemize}
  \item the derivative of $(\ref{10})$ with respect to $\hat{\lambda}_k$ is equal to the
  derivative of $(\ref{13})$ with respect to $\lambda$, thanks to $(\ref{11})$, $(\ref{15})_1$,
  \item the derivative of $(\ref{10})$ with respect to $\hat{\lambda}_{kb}$ is exactly the
  derivative of $(\ref{13})$ with respect to $\lambda_b$, thanks to $(\ref{11})$, $(\ref{15})_{2,1}$,
  \item the derivative of $(\ref{10})$ with respect to $\hat{\lambda}_{kll}$ is exactly the
  derivative of $(\ref{13})$ with respect to $\lambda_{ab}$, contracted after derivation by $\delta_{ab}$,
  thanks to $(\ref{11})$, $(\ref{15})_{3,2}$,
  \item the derivative of $(\ref{10})$ with respect to $\hat{\lambda}_{kkll}$ is exactly the
  derivative of $(\ref{13})$ with respect to $\lambda_{ill}$, contracted after derivation by $\delta_{ki}$,
  thanks to $(\ref{11})$, $(\ref{15})_{5}$.
\end{itemize}
Consequently, eq. $(\ref{10})$ needs to be imposed only for $\hat{\lambda}_k=0$,
$\hat{\lambda}_{ab}=0$, $\hat{\lambda}_{kll}=0$ and $\hat{\lambda}_{kkll}=0$, and in this
case it is an identity. So it remains to impose only eq. $(\ref{13})$. To this end it is
useful to use the identity
\begin{eqnarray*}
  \frac{\partial^r}{\partial \hat{\lambda}_{k_1h_1}\cdots \partial \hat{\lambda}_{k_r
  h_r}}\left(\hat{\lambda}_{ij}\frac{\partial \phi^{'k}}{\partial
  \hat{\lambda}_{i}}\right)&=&\hat{\lambda}_{ij}\frac{\partial^{r+1} \phi^{'k}}{\partial \hat{\lambda}_{i}\partial
\hat{\lambda}_{k_1h_1}\cdots\partial
  \hat{\lambda}_{k_rh_r}}\\
  &+&r\delta_{j(k_1}\frac{\partial^{r} \phi^{'k}}{\partial \hat{\lambda}_{h_1}\partial \hat{\lambda}_{k_2h_2}\cdots\partial
  \hat{\lambda}_{k_rh_r)}}
\end{eqnarray*}
whose proof can be found in the Appendix of ref. [5] and holds also if, in our case,
$\hat{\phi}^{'k}$ depends on the further independent variable $\hat{\lambda}_{aabb}$.\\
Let us take now the derivative of eq. $(\ref{13})$ with res\-pect to
$\hat{\lambda}_{i_1}\cdots\hat{\lambda}_{i_p}$,
$\hat{\lambda}_{j_1ll}\cdots\hat{\lambda}_{j_qll}$,
$\hat{\lambda}_{k_1h_1}\cdots\hat{\lambda}_{k_rh_r}$. If we calculate it at equilibrium
and we use eqs. $(\ref{17})$ and $(\ref{20})$ we obtain
\begin{eqnarray}\label{e1}
&& p \delta_{h(i_1}\frac{\partial}{\partial \hat{\lambda}}\phi_{p-1,q,r}^{i_2\cdots
i_p)kj_1\cdots j_qh_1k_1\cdots h_rk_r }+\frac{2}{3}\hat{\lambda}_{ll}\phi_{p+1,q,r}^{kh
i_1\cdots i_pj_1\cdots j_qh_1k_1\cdots h_rk_r
}+\nonumber \\
&+&2r\delta_{h(k_1}\phi_{p+1,q,r-1}^{h_1k_2\cdots h_rk_r )ki_1\cdots i_p j_1\cdots
j_q}+2q\phi_{p,q-1,r+1}^{khi_1\cdots i_pj_1\cdots j_qh_1k_1\cdots h_rk_r }+\nonumber \\
&+&q\delta_{h(j_1}\phi_{p,q-1,r+1}^{j_2\cdots j_q)ki_1\cdots i_ph_1k_1\cdots h_rk_r
ab}\delta_{ab}+ 4\hat{\lambda}_{aabb}\phi_{p,q+1,r}^{ki_1\cdots i_pj_1\cdots
j_qhh_1k_1\cdots h_rk_r}+\nonumber \\
&+& h_{p,q,r}^{i_1\cdots i_pj_1\cdots j_qh_1k_1\cdots h_rk_r }\delta^{hk}=0.
\end{eqnarray}
To evaluate this condition it will be useful to do the following considerations: \\~~~\\
1) Let $\psi^{\cdots}$ be a symmetric tensor; it is easy to prove that
\begin{eqnarray*}
\delta^{h(i_1}\psi^{i_2\cdots i_p j_1\cdots j_q e_1\cdots e_s
k)}&=&\frac{p}{p+q+s+1}\delta^{h(i_1}\psi^{i_2\cdots i_p )j_1\cdots j_q e_1\cdots e_s k}+\\
&+&\frac{q}{p+q+s+1}\delta^{h(j_1}\psi^{j_2\cdots j_q)i_1\cdots i_p e_1\cdots e_s k}+\\
&+&\frac{s}{p+q+s+1}\delta^{h(e_1}\psi^{e_2\cdots e_s)i_1\cdots i_p j_1\cdots j_q
k}+\\&+&\frac{1}{p+q+s+1}\delta^{hk}\psi^{i_1\cdots i_p j_1\cdots j_q e_1\cdots e_s}.
\end{eqnarray*}
2) Moreover we have
\begin{eqnarray*}
\phi_{p,q-1,r+1}^{j_2\cdots j_qki_1\cdots i_ph_1k_1\cdots h_rk_r
ab}\delta_{ab}=\phi_{p,q-1,r+1}\frac{q+p+2r+3}{q+p+2r+1}\delta^{(j_2\cdots j_qki_1\cdots
i_ph_1k_1\cdots h_rk_r )}.
\end{eqnarray*}
3) Finally, we can express everything in terms of the scalar $h_{p,q,r}$ using the
following relations:
\begin{equation*}
  \begin{cases}
    \frac{\partial }{\partial \hat{\lambda}}\phi_{p-1,q,r}=h_{p,q,r} & \text{from eq. $(\ref{alfa})$}, \\
    \phi_{p+1,q,r-1}=h_{p,q,r},\quad \phi_{p+1,q,r}=h_{p,q,r+1} & \text{from eq. $(\ref{gamma})$}, \\
    \phi_{p,q-1,r+1}=\frac{p+q+2r+1}{p+q+2r+3}h_{p,q,r} & \text{from eq. $(\ref{delta})$}, \\
    \phi_{p,q+1,r}=\frac{p+q+2r+1}{p+q+2r+3}\frac{\partial}{\partial \hat{\lambda}_{aabb}}h_{p,q,r} & \text{from eq.
    $(\ref{epsilon})$}.
  \end{cases}
\end{equation*}
All these results allow to rewrite eq. $(\ref{e1})$ as
\begin{eqnarray*}
0=h_{p,q,r}(p+q+2r+1)\delta^{h(i_1}\delta^{i_2\cdots
h_rk_rk)}+\frac{2}{3}\hat{\lambda}_{ll}
\delta^{kh i_1\cdots h_rk_r}h_{p,q,r+1}+\\
+2q \frac{p+q+2r+1}{p+q+2r+3}\delta^{kh i_1\cdots
h_rk_r}h_{p,q,r}+4\hat{\lambda}_{aabb}\frac{p+q+2r+1}{p+q+2r+3}\frac{\partial
h_{p,q,r}}{\partial \hat{\lambda}_{aabb}}\delta^{hk i_1\cdots h_rk_r},
\end{eqnarray*}
where the notation $\delta^{e_1e_2\cdots e_{2s}}=\delta^{(e_1e_2}\cdots
\delta^{e_{2s-1}e_{2s})}$ has been used; the result is equivalent to
\begin{eqnarray*}
 0&=&(p+q+2r+1)h_{p,q,r}+\frac{2}{3}\hat{\lambda}_{ll}
h_{p,q,r+1}+\\&+& \frac{p+q+2r+1}{p+q+2r+3}\left(2q
h_{p,q,r}+4\hat{\lambda}_{aabb}\frac{\partial h_{p,q,r}}{\partial
\hat{\lambda}_{aabb}}\right).
\end{eqnarray*}
This equation, by using eqs. $(\ref{25})$ and $(\ref{definition})$, becomes
\begin{eqnarray*}
 0=(p+3q+2r+3)\frac{\partial^r}{\partial \hat{\lambda}^r_{ll}}k_{p,q}+2\hat{\lambda}_{ll}
\frac{\partial^{r+1}}{\partial \hat{\lambda}^{r+1}_{ll}}k_{p,q}+
4\hat{\lambda}_{aabb}\frac{\partial^r}{\partial \hat{\lambda}^r_{ll}}\frac{\partial
k_{p,q}}{\partial \hat{\lambda}_{aabb}}
\end{eqnarray*}
with p+q even. We note that if this relation holds until a fixed r taking its derivative
with respect to $\hat{\lambda}_{ll}$ we obtain that it holds also with r+1 replacing r.
Therefore, it suffices to impose this relation for the lower value of r, i.e for r=0. In
this case it becomes
\begin{eqnarray}\label{eq}
 0=(p+3q+3)k_{p,q}+2\hat{\lambda}_{ll}
\frac{\partial}{\partial \hat{\lambda}_{ll}}k_{p,q}+ 4\hat{\lambda}_{aabb}\frac{\partial
k_{p,q}}{\partial \hat{\lambda}_{aabb}},
\end{eqnarray}
with p+q even.\\
Let us firstly analyze the case with p and q even. Putting eq. $(\ref{betagamma})_1$ into
$(\ref{eq})$ we have
\begin{eqnarray*}
 0=(p+3q+3)\frac{\partial^p}{\partial \hat{\lambda}^{\frac{p}{2}}_{ll}\partial \hat{\lambda}^{\frac{p}{2}}}k_{0,q}
 +2\hat{\lambda}_{ll}\frac{\partial^{p+1}}{\partial \hat{\lambda}^{\frac{p}{2}+1}_{ll}\partial
\hat{\lambda}^{\frac{p}{2}}}k_{0,q}+ 4\hat{\lambda}_{aabb}\frac{\partial^p}{\partial
\hat{\lambda}^{\frac{p}{2}}_{ll}\partial \hat{\lambda}^{\frac{p}{2}}}\frac{\partial
k_{0,q}}{\partial \hat{\lambda}_{aabb}}.
\end{eqnarray*}
We note that if this relation holds until a fixed p taking its derivative with respect to
$\hat{\lambda}_{ll}$ and then with respect to $\hat{\lambda}$, we obtain that it holds
also with p+2 replacing p (p must be even). Therefore, it suffices to impose this
relation for the lower even value of p, i.e for p=0. \\ In this case it becomes
\begin{eqnarray}\label{eq1}
 0=(3q+3)k_{0,q}+2\hat{\lambda}_{ll}
\frac{\partial}{\partial \hat{\lambda}_{ll}}k_{0,q}+ 4\hat{\lambda}_{aabb}\frac{\partial
k_{0,q}}{\partial \hat{\lambda}_{aabb}},
\end{eqnarray}
that is $(\ref{eq})$ calculated in p=0.\\
By using eq. $(\ref{A})$ we see that eq. $(\ref{eq1})$ becomes
\begin{eqnarray*}
 0=(3q+3)\frac{\partial^q}{\partial \hat{\lambda}^{\frac{q}{2}}_{ll}\partial \hat{\lambda}_{aabb}^{\frac{q}{2}}}k_{0,0}
 +2\hat{\lambda}_{ll}\frac{\partial^{q+1}}{\partial \hat{\lambda}^{\frac{q}{2}+1}_{ll}\partial
\hat{\lambda}_{aabb}^{\frac{q}{2}}}k_{0,0}+
4\hat{\lambda}_{aabb}\frac{\partial^q}{\partial \hat{\lambda}^{\frac{q}{2}}_{ll}\partial
\hat{\lambda}_{aabb}^{\frac{q}{2}}}\frac{\partial k_{0,0}}{\partial
\hat{\lambda}_{ccgg}}.
\end{eqnarray*}
We note that if this relation holds until a fixed q taking its derivative with respect to
$\hat{\lambda}_{ll}$ and then with respect to $\hat{\lambda}_{aabb}$, we obtain that it
holds also with q+2 replacing q (q must be even). Therefore, it suffices to impose this
relation for the lower even order of q, i.e for q=0. In this case it becomes
\begin{eqnarray}\label{f1}
 0=3k_{0,0}+2\hat{\lambda}_{ll}
\frac{\partial}{\partial \hat{\lambda}_{ll}}k_{0,0}+ 4\hat{\lambda}_{aabb}\frac{\partial
k_{0,0}}{\partial \hat{\lambda}_{aabb}},
\end{eqnarray}
that is $(\ref{eq})$ calculated in p=0, q=0.\\
There remains the case with p and q odd. We will see that it will give only identities.
In fact, putting eq. $(\ref{betagamma})_2$ into $(\ref{eq})$, this becomes
\begin{eqnarray*}
 0&=&(p+3q+3)\frac{\partial^p}{\partial \hat{\lambda}^{\frac{p-1}{2}}_{ll}\partial \hat{\lambda}^{\frac{p+1}{2}}}k_{0,q}
 +\\&+&2\hat{\lambda}_{ll}\frac{\partial^{p+1}}{\partial \hat{\lambda}^{\frac{p-1}{2}+1}_{ll}\partial
\hat{\lambda}^{\frac{p+1}{2}}}k_{0,q}+ 4\hat{\lambda}_{aabb}\frac{\partial^p}{\partial
\hat{\lambda}^{\frac{p-1}{2}}_{ll}\partial \hat{\lambda}^{\frac{p+1}{2}}}\frac{\partial
k_{0,q}}{\partial \hat{\lambda}_{aabb}}.
\end{eqnarray*}
We note that if this relation holds until a fixed p taking its derivative with respect to
$\hat{\lambda}_{ll}$ and then with respect to $\hat{\lambda}$, we obtain that it holds
also with p+2 replacing p (p must be odd). Therefore, it suffices to impose this relation
for the lower odd value of p, i.e p=1. In this case it becomes
\begin{eqnarray}\label{eq2}
 0=(3q+4)\frac{\partial}{\partial \hat{\lambda}}k_{0,q}+2\hat{\lambda}_{ll}
\frac{\partial^2}{\partial \hat{\lambda}\partial\hat{\lambda}_{ll}}k_{0,q}+
4\hat{\lambda}_{aabb}\frac{\partial^2 k_{0,q}}{\partial\hat{\lambda}\partial
\hat{\lambda}_{aabb}}.
\end{eqnarray}
This relation, by using eq. $(\ref{B})$ becomes
\begin{eqnarray*}
 0&=&(3q+4)\frac{\partial^q}{\partial \hat{\lambda}^{\frac{q-1}{2}}_{ll}\partial \hat{\lambda}_{aabb}^{\frac{q+1}{2}}}
 \frac{\partial}{\partial\hat{\lambda}}k_{0,0}+\\
&+&2\hat{\lambda}_{ll}\frac{\partial^{q+1}}{\partial
\hat{\lambda}^{\frac{q-1}{2}+1}_{ll}\partial
\hat{\lambda}_{aabb}^{\frac{q+1}{2}}}\frac{\partial}{\partial\hat{\lambda}}k_{0,0}+
4\hat{\lambda}_{aabb}\frac{\partial^q}{\partial
\hat{\lambda}^{\frac{q-1}{2}}_{ll}\partial
\hat{\lambda}_{aabb}^{\frac{q+1}{2}}}\frac{\partial^2 k_{0,0}}{\partial
\hat{\lambda}_{ccgg}\partial\hat{\lambda}}.
\end{eqnarray*}
We note that if this relation holds until a fixed q, taking its derivative with respect
to $\hat{\lambda}_{ll}$ and then with respect to $\hat{\lambda}_{aabb}$, we obtain that
it holds also with q+2 replacing q (q must be odd). Therefore, it suffices to impose this
relation for the lower odd value of q, i.e q=1. In this case it becomes
\begin{eqnarray*}
 0=7\frac{\partial^2}{\partial \hat{\lambda}\partial\hat{\lambda}_{ppqq}}k_{0,0}+2\hat{\lambda}_{ll}
\frac{\partial^3}{\partial
\hat{\lambda}\partial\hat{\lambda}_{ll}\partial\hat{\lambda}_{ppqq}}k_{0,0}+
4\hat{\lambda}_{ppqq}\frac{\partial^3 k_{0,0}}{\partial\hat{\lambda}\partial
\hat{\lambda}^2_{ppqq}},
\end{eqnarray*}
which is a consequence of $(\ref{f1})$ because it is its second derivative with respect
to $\hat{\lambda}$ and $\hat{\lambda}_{ppqq}$. In this way, we have seen that the
conditions $(\ref{10})$
 and $(\ref{13})$ give only the restriction $(\ref{f1})$ for $k_{0,0}$ and many identities.\\
\textbf{So we have that every element of the matrix $k_{p,q}$ can be expressed as
function of $k_{0,0}$ and this is restricted only by eqs. $(\ref{C})$ and $(\ref{f1})$.}\\
Let us conclude by exploiting these conditions and let us do it by using the expansion of
$k_{0,0}$ around the state with $\hat{\lambda}_{ppqq}=0$, i.e.,
\begin{equation}\label{D}
k_{0,0}=\sum_{s=0}^{\infty}\frac{1}{s!}k_s(\hat{\lambda},\hat{\lambda}_{ll})\hat{\lambda}_{ppqq}^s.
\end{equation}
Using $(\ref{D})$, eq. $(\ref{C})$ becomes
\begin{equation}\label{E}
9\frac{\partial^2 k_s}{\partial \hat{\lambda}^2_{ll}}=\frac{\partial k_{s+1}}{\partial
\hat{\lambda}},
\end{equation}
while eq. $(\ref{f1})$ transforms into
\begin{equation*}
0=3\sum_{s=0}^{\infty}\frac{1}{s!}k_s \hat{\lambda}_{ppqq}^s
+2\hat{\lambda}_{ll}\sum_{s=0}^{\infty}\frac{1}{s!}\frac{\partial k_s}{\partial
\hat{\lambda}_{ll}}
\hat{\lambda}_{ppqq}^s+4\sum_{s=1}^{\infty}\frac{1}{(s-1)!}k_s\hat{\lambda}_{ppqq}^s
\end{equation*}
\begin{equation*}
\text{i.e. ~~~\qquad \qquad }  \begin{cases}
    3k_0+2\hat{\lambda}_{ll}\frac{\partial k_0}{\partial \hat{\lambda}_{ll}}=0 & \text{for s=0}, \\
    3k_s+2\hat{\lambda}_{ll}\frac{\partial k_s}{\partial \hat{\lambda}_{ll}}+4sk_s=0 & \text{for $s\geq
    1$};\qquad \qquad \qquad \qquad \qquad
  \end{cases}
\end{equation*}
but the relation for s=0 is contained in the other equation, so that they can be written
as
\begin{equation*}
(3+4s)k_s+2\hat{\lambda}_{ll}\frac{\partial k_s}{\partial\hat{\lambda}_{ll}}=0
\quad\forall s\geq 0,
\end{equation*}
whose solution is
\begin{equation}\label{h1}
k_s=\hat{\lambda}_{ll}^{-\frac{3+4s}{2}}\widetilde{k}_s(\hat{\lambda}).
\end{equation}
This allows to rewrite eq. $(\ref{E})$ as
\begin{equation}\label{49bis}
\frac{\partial \widetilde{k}_{s+1}}{\partial
\hat{\lambda}}=\widetilde{k}_s\frac{9}{4}(3+4s)(5+4s).
\end{equation}
In this way we have found that $\widetilde{k}_0(\hat{\lambda})$ is an arbitrary
single-variable function, while the other functions $\widetilde{k}_{s+1}(\hat{\lambda})$
are determined by $(\ref{49bis})$, except for a numerable family of constants arising
from integration.
\begin{center} \textbf{5. The 13 moments model as a subsystem of the 14 moments one} \end{center}
To verify that the 13 moments case is a subsystem of the 14 moments one we will show that
the relations obtained in [5] for the scalar functions $j_{0,q}$ are satisfied by the
value of $k_{0,q}$ found here but considering $\hat{\lambda}_{ppqq}=0$. Firstly we have
to rewrite the expressions of $k_{0,q}$. Substituting eq. $(\ref{D})$ into eq.
$(\ref{A})$ we have
\begin{eqnarray*}
k_{0,q}&=&3^{\frac{q}{2}}\frac{1}{q+1}\sum_{s=0}^{\infty}\frac{1}{s!}\frac{\partial^{\frac{q}{2}}k_s}
{\partial\hat{\lambda}_{ll}^{\frac{q}{2}}}\frac{\partial^{\frac{q}{2}}\hat{\lambda}_{ppqq}^{s}}{\partial\hat{\lambda}_{ppqq}^{\frac{q}{2}}}=\\
&=&3^{\frac{q}{2}}\frac{1}{q+1}\sum_{s=\frac{q}{2}}^{\infty}\frac{1}{s!}\frac{\partial^{\frac{q}{2}}k_s}
{\partial\hat{\lambda}_{ll}^{\frac{q}{2}}}s(s-1)\cdots (s-\frac{q}{2}+1)
\hat{\lambda}_{ppqq}^{s-\frac{q}{2}}
\end{eqnarray*}
If we calculate this for $\hat{\lambda}_{ppqq}=0$, only the term for $s=\frac{q}{2}$
remains, so our relations becomes
\begin{eqnarray*}
k_{0,q}=3^{\frac{q}{2}}\frac{1}{q+1}\frac{\partial^{\frac{q}{2}}k_{\frac{q}{2}}}
{\partial\hat{\lambda}_{ll}^{\frac{q}{2}}}\quad \text{with q even.}
\end{eqnarray*}
Substituting eq. $(\ref{D})$ into eq. $(\ref{B})$, still making the previous
considerations, we have
\begin{eqnarray*}
k_{0,q}=3^{\frac{q-1}{2}}\frac{1}{q+2}\frac{\partial^{\frac{q-1}{2}}k_{\frac{q+1}{2}}}
{\partial\hat{\lambda}_{ll}^{\frac{q-1}{2}}}\quad \text{with q odd.}
\end{eqnarray*}
Now using eqs. $(\ref{h1})$ we obtain
\begin{equation}\label{df}
k_{0,q}= \begin{cases}
    3^{\frac{q}{2}}\frac{1}{q+1}\left(-\frac{1}{2}\right)^{\frac{q}{2}}\eta(3+2q,3q+1)\hat{\lambda}_{ll}^{-\frac{3+3q}{2}}\tilde{k}_{\frac{q}{2}}& \text{for q even}, \\
    \\
    3^{\frac{q-1}{2}}\frac{1}{q+2}\left(-\frac{1}{2}\right)^{\frac{q-1}{2}}\eta(5+2q,3q+2)\hat{\lambda}_{ll}^{-\frac{4+3q}{2}}\tilde{k}_{\frac{q+1}{2}} & \text{for q odd}.
  \end{cases}
\end{equation}
where $\eta(a,b)=a(a-2)(a-4)\cdots(b+2)b$.\\
Comparing this result with the corresponding one for $j_{0,q}$ in [5] (i.e. eqs. (56) and
(57)), we find that they are the same, except for identifying
\begin{equation}\label{51}
I_q(\hat{\lambda})=\left(-\frac{3}{2}\right)^{\frac{q}{2}}\frac{1}{q+1}\eta(3+2q,3q+1)\widetilde{k}_{\frac{q}{2}}(\hat{\lambda})
\end{equation}
and for setting $c_q=0$.\\
It is easy to verify that with $I_q$ given by eq. $(\ref{51})$, the condition (58) of
ref. [5] becomes exactly the present eq. $(\ref{49bis})$, except for substituting q=2s+2,
and viceversa. All the other results of ref. [5], for the 13 moments model, can be
obtained by substituting $\hat{\lambda}_{aabb}=0$ in the present ones except for
the new restriction $c_q=0$.\\
In other words, in ref. [5] the solution was found except for two families of constants,
one arising from integration of eq. $(58)$ in ref. [5] and another constituted by the
constants $c_q$ appearing in eq. (57). This second family of constants doesn't appear if
the 13 moments model is obtained as a subsystem of the 14 moments one.
\begin{center} \textbf{6. The comparison with the kinetic approach} \end{center}
The solution of our conditions proposed by the kinetic approach, see [2] and [9], is
\begin{eqnarray*}
h'&=&\int F(\lambda+\lambda_i c^i+\lambda_{ij}c^i c^j+\lambda_{ill}c^i
c^2+\lambda_{aabb}c^4)dc_1dc_2dc_3\\
\phi^{'k}&=&\int F(\lambda+\lambda_i c^i+\lambda_{ij}c^i c^j+\lambda_{ill}c^i
c^2+\lambda_{aabb}c^4)c^kdc_1dc_2dc_3,
\end{eqnarray*}
(where F is related with the distribution function at equilibrium), and it is easy to see
that it satisfies the conditions $(\ref{15})$, $(\ref{10})$, $(\ref{13})$. We can now see
that it is a particular case of our general solution. In fact eqs. $(\ref{17})$ and
$(\ref{20})$ now become
\begin{eqnarray*}
\phi_{p,q,r}^{ii_1\cdots i_pj_1\cdots j_q k_1h_1\cdots k_rh_r}&=&\int
F^{(p+q+r)}(\lambda+\frac{1}{3}\lambda_{ll}c^2+\lambda_{aabb}c^4)\\
&&c^ic^{i_1}\cdots c^{i_p}c^{j_1}\cdots c^{j_q}c^{2q}c^{h_1}c^{k_1}\cdots c^{h_r}c^{k_r}dc_1dc_2dc_3,\\
h_{p,q,r}^{i_1\cdots i_pj_1\cdots j_q k_1h_1\cdots k_rh_r}&=&\int
F^{(p+q+r)}(\lambda+\frac{1}{3}\lambda_{ll}c^2+\lambda_{aabb}c^4)\\
&&c^{i_1}\cdots c^{i_p}c^{j_1}\cdots c^{j_q}c^{2q}c^{h_1}c^{k_1}\cdots
c^{h_r}c^{k_r}dc_1dc_2dc_3,
\end{eqnarray*}
and it is easy to see that eqs. $(\ref{22})$ and $(\ref{23})$ are satisfied.\\
Eqs. $(\ref{18})$ and $(\ref{21})$ hold with
\begin{eqnarray*}
\phi_{p,q,r}=\frac{4\pi}{p+q+2r+2}\int_{0}^{\infty}
F^{(p+q+r)}(\lambda+\frac{1}{3}\lambda_{ll}c^2+\lambda_{aabb}c^4)c^{p+3q+2r+3}dc,\\
h_{p,q,r}=\frac{4\pi}{p+q+2r+1}\int_{0}^{\infty}
F^{(p+q+r)}(\lambda+\frac{1}{3}\lambda_{ll}c^2+\lambda_{aabb}c^4)c^{p+3q+2r+2}dc.
\end{eqnarray*}
The eqs. $(\ref{25})$ and $(\ref{27})$ are consequences of these. The definitions
$(\ref{definition})$ now become
\begin{eqnarray*}
k_{p,q}&=&\frac{4\pi}{p+q+1}\int_{0}^{\infty}
F^{(p+q)}(\lambda+\frac{1}{3}\lambda_{ll}c^2+\lambda_{aabb}c^4)c^{p+3q+2}dc \text{if p+q is even},\\
k_{p,q}&=&\frac{4\pi}{p+q+2}\int_{0}^{\infty}
F^{(p+q)}(\lambda+\frac{1}{3}\lambda_{ll}c^2+\lambda_{aabb}c^4)c^{p+3q+3}dc \text{if p+q
is odd}.
\end{eqnarray*}
From these it follows
\begin{equation*}
k_{0,0}=4\pi\int_{0}^{\infty}
F(\lambda+\frac{1}{3}\lambda_{ll}c^2+\lambda_{aabb}c^4)c^{2}dc
\end{equation*}
and it is not difficult to see that eqs. $(\ref{C})$ and $(\ref{40bis})$ are satisfied.\\
Proof of eq. $(\ref{f1})$ needs an integration by parts, as follows
\begin{eqnarray*}
0&=&3\int_{0}^{\infty}Fc^2 dc
+\frac{2}{3}\lambda_{ll}\int_{0}^{\infty}F'c^4dc+4\lambda_{aabb}\int_{0}^{\infty}F'c^6dc=\\
&=&3\int_{0}^{\infty}Fc^2 dc +\int_{0}^{\infty}\left(\frac{dF}{dc}\right)c^3dc=
3\int_{0}^{\infty}Fc^2 dc + \left|Fc^3 \right|_{0}^{\infty}-\int_{0}^{\infty}3Fc^2dc
\end{eqnarray*}
which is satisfied because
\begin{equation*}
\lim_{c\rightarrow\infty}Fc^3=0.
\end{equation*}
We can now see that eq. $(\ref{D})$ holds with
\begin{equation*}
k_{s}=4\pi\int_{0}^{\infty} F^{(s)}(\lambda+\frac{1}{3}\lambda_{ll}c^2)c^{4s+2}dc,
\end{equation*}
of which eq. $(\ref{E})$ is an easy consequence.\\
By using the change of the integration variables $c=\eta\lambda_{ll}^{-\frac{1}{2}}$, we
obtain eq. $(\ref{h1})$ with
\begin{equation}\label{52}
\widetilde{k}_{s}=4\pi\int_{0}^{\infty}
F^{(s)}(\lambda+\frac{1}{3}\eta^2)\eta^{4s+2}d\eta.
\end{equation}
Proof of eq. $(\ref{h1})$ needs two integrations by part, as follows
\begin{eqnarray*}
\frac{d}{d\lambda}\widetilde{k}_{s+1}&=&4\pi\int_{0}^{\infty}
F^{(s+2)}(\lambda+\frac{1}{3}\eta^2)\eta^{4s+6}d\eta=\\
&=&\left|4\pi
F^{(s+1)}(\lambda+\frac{1}{3}\eta^2)\frac{3}{2}\eta^{4s+5}\right|_{0}^{\infty}+
\\&-&\int_{0}^{\infty}6\pi(4s+5)F^{(s+1)}(\lambda+\frac{1}{3}\eta^2)\eta^{4s+4}d\eta=\\
&=&\left|-6\pi(4s+5)F^{(s)}(\lambda+\frac{1}{3}\eta^2)\frac{3}{2}\eta^{4s+3}\right|_{0}^{\infty}
+\\
&-&\int_{0}^{\infty}- 9\pi(4s+5)(4s+3)F^{(s)}(\lambda+\frac{1}{3}\eta^2)\eta^{4s+2}d\eta
=\\
&=&\frac{9}{4}(4s+3)(4s+5)\widetilde{k}_{s}.
\end{eqnarray*}
Consequently, the kinetic approach suggest to take
\begin{equation*}
\widetilde{k}_{0}(\lambda)=4\pi\int_{0}^{\infty}
F(\lambda+\frac{1}{3}\eta^2)\eta^{2}d\eta,
\end{equation*}
which is only a change from our arbitrary function $\widetilde{k}_{0}(\lambda)$ to the
arbitrary function F; moreover it considers only a particular solution of the eqs.
$(\ref{49bis})$, i.e., eq. $(\ref{52})$. In this way the numerable family of arbitrary
constants arising from integration of eq. $(\ref{49bis})$ doesn't appear in the kinetic
approach. Then the macroscopic approach here considered is more general than the kinetic
one.\newpage
\begin{center} \textbf{7. Conclusions} \end{center}
We are very satisfied by the present results because we have found the closure of the
field equations up to whatever order with respect to equilibrium. This was never obtained
before in literature. Apparently difficulties in calculation becomes very elegant and
somehow simpler. Moreover, our model inherits all the nice properties of the Extended
Thermodynamics, such as to be expressed in the form of a symmetric hyperbolic system, to
predict finite speeds of propagation for shock waves, and to guarantee well posedness of
the Cauchy problem and continuous dependence on the initial data.
\begin{center} \textbf{References} \end{center}
[1] G.M. Kremer, Extended Thermodynamics of ideal gases with 14
fields. Ann Inst. Henri Poincar$\acute{e}$, 45, p.419,
(1986),\newline\newline [2] I. M\"{u}ller, T. Ruggeri, Rational
Extended Thermodynamics, second edition. Springer-Verlag, New
York, Berlin Heidelberg (1998),\newline\newline [3] I-Shih Liu,
Method of Lagrange multipliers for exploitation of the entropy
principle. Arch. Rational. Mech. Anal. \underline{46}, p.131
(1972),\newline\newline [4] T. Ruggeri, A. Strumia, Main field and
convex covariant density for quasi-linear hyperbolic systems,
Relativistic fluid dynamics, Ann. Inst. H. Poincarr$\grave{e}$,
\underline{34}, p.65 (1981),\newline\newline [5] S. Pennisi, T.
Ruggeri, A new method to exploit the entropy principle and
galilean invariance in the macroscopic approach of extended
thermodynamics, Ricerche di Matematica, Springer, \underline{55},
p.319 (2006), \newline\newline [6] S. Pennisi, A. Scanu, Judicious
interpretation of the condition present in extended
thermodynamics. Proceedings of Wascom 2003, Villasimius (Cagliari)
1-7.6.2005. Word Scientific, Singapore, p.393
(2003),\newline\newline [7] T. Ruggeri, Galilean invariance and
entropy principle for systems of balance laws, Continuum Mech.
Thermodyn., \underline{1}, p.3, (1989),\newline\newline [8] G.
Boillat, T. Ruggeri, Hyperbolic principal subsystems: Entropy
convexity and sub characteristics conditions, Arch. Rat. Mech.
Anal. \underline{137}, p.303, (1997),\newline\newline [9] G.
Boillat, T. Ruggeri, Moment equations in the kinetic theory of
gases and wave velocities, Continuum Mech. Thermodyn.
\underline{9}, p.205, (1997).
\end{document}